\documentclass[conference,letterpaper]{IEEEtran}

\usepackage[utf8]{inputenc} 
\usepackage[T1]{fontenc}
\usepackage{url}
\usepackage{ifthen}
\usepackage{cite}
\usepackage[cmex10]{amsmath}
\usepackage{amssymb}
\usepackage{graphicx}

\newtheorem{thm}{Theorem}
\newtheorem{rk}{Remark}
\newtheorem{cor}{Corollary}

\usepackage{flushend} 
\bstctlcite{IEEEexample:BSTcontrol} 
\usepackage[all=normal, paragraphs=tight, floats=tight, mathspacing=tight]{savetrees}

\begin{document}
\title{Detection of Signals in Colored Noise: Leading Eigenvalue Test for Non-central $F$-matrices
\vspace{-4mm}
} 



\author{%
  \IEEEauthorblockN{Prathapasinghe~Dharmawansa\IEEEauthorrefmark{1},
                    Saman Atapattu\IEEEauthorrefmark{2},
                    Jamie Evans\IEEEauthorrefmark{1},
                    and Kandeepan Sithamparanathan\IEEEauthorrefmark{2}}
  \IEEEauthorblockA{\IEEEauthorrefmark{1}%
                   Department of Electrical and Electronic Engineering, University of Melbourne, Australia,
                    \{prathapa.d, jse\}@unimelb.edu.au}
  \IEEEauthorblockA{\IEEEauthorrefmark{2}%
                    School of Engineering, RMIT University, Australia,
                    \{saman.atapattu, kandeepan.sithamparanathan\}@rmit.edu.au}
\vspace{-8mm}}

\maketitle


\begin{abstract}
   This paper investigates the signal detection problem in colored noise with an unknown covariance matrix. In particular, we focus on detecting an unknown non-random signal by capitalizing on the leading eigenvalue of the whitened sample covariance matrix as the test statistic (a.k.a. Roy's largest root test). Since the unknown signal is non-random, the whitened sample covariance matrix turns out to have a {\it non-central} $F$-distribution. This distribution assumes a singular or non-singular form depending on whether the number of observations $p\lessgtr$ the system dimensionality $m$. Therefore, we statistically characterize the leading eigenvalue of the singular and non-singular $F$-matrices by deriving their cumulative distribution functions (c.d.f.). Subsequently, they have been utilized in deriving the corresponding receiver operating characteristic (ROC) profiles. We also extend our analysis into the high dimensional domain. It turns out that, when the signal is sufficiently strong, the maximum eigenvalue can reliably detect it in this regime. Nevertheless, {\it weak signals} cannot be detected in the high dimensional regime with the leading eigenvalue. 
\end{abstract}

\section{Introduction}
The fundamental problem of detecting a signal embedded in noise has been in the forefront of various research studies, see e.g., \cite{nadakuditi,chamain2020eigenvalue,debbah2011,edelman,mcwhorter2023,johnstone2020,YCliang,larson,hemimo} and references therein. In this regard, among various test statistics available, the leading eigenvalue (a.k.a. Roy's largest root) of the sample covariance matrix has been a popular choice among the detection theorists \cite{Kritchman,johnstoneRoy,chamain2020eigenvalue,prathapRoyRoot,wang2017stat,YCliang,larson}. This stems from the fact that the largest
root test is most powerful among the common tests when
the alternative is of rank-one \cite{johnstoneRoy,Kritchman}. The rank-one alternative manifests as a single {\it spike} with respect to either the covariance or non-centrality parameter matrices \cite{johnstone2001,baikPhase,johnstoneRoy,Dharmawansa2014}.

The additive white Gaussian noise assumption, though very common in the traditional setting, may not hold in certain modern practical scenarios \cite{richmond,hemimo,mcwhorter2023,Vinogradova,werner2007,werner2006}. In such situations, the detection theorists prefer to work with respect to a transformed and re-scaled coordinate system in which the effective noise is white \cite{chamain2020eigenvalue,nadakuditi,johnstoneRoy,muirhead}. This operation is commonly known as the {\it whitening}. In this respect, the unknown noise covariance matrix is usually estimated with the so called {\it noise-only} (i.e., signal free or secondary data) samples. The availability of noise-only samples is a commonly  made assumption in the literature \cite{dogandzic2003generalized,nadakuditi,johnstoneRoy,richmond,werner2007,zachariah,melvin2004}.

Assuming the availability of primary (i.e., plausible signal-plus-noise samples)  and secondary data sets, the whitened sample covariance matrix can conveniently be written as $\widehat{\boldsymbol{F}}=\widehat{\boldsymbol{\Sigma}}^{-1/2}\boldsymbol{\widehat R}\widehat{\boldsymbol{\Sigma}}^{-1/2}$ where  $\widehat{\boldsymbol{\Sigma}}\in\mathbb{C}^{m\times m}$ is the estimated noise covariance matrix, $\widehat{\boldsymbol{R}}\in\mathbb{C}^{m\times m}$ denotes the sample covariance estimated with the primary data, $m$ is the system dimensionality and $(\cdot)^{1/2}$ denotes the positive definite square root. Since the signal of our interest is unknown but non-random, under the Gaussian noise assumption, $\widehat{\boldsymbol{R}}\in\mathbb{C}^{m\times m}$ is non-central Wishart distributed, whereas $\widehat{\boldsymbol{\Sigma}}\in\mathbb{C}^{m\times m}$ has a central Wishart distribution. Consequently,  $\widehat{\boldsymbol{F}}$ follows  the so-called non-central $F$-distribution \cite{james,muirhead,johnstoneRoy}. In certain situations, the number of available samples may be little in comparison to the system dimension $m$ (i.e., sample deficiency) \cite{Vinogradova,vallet2017performance,pham2015,mestre2008}. In such situation, the matrix $\widehat{\boldsymbol{F}}$ becomes rank deficient and there by having a singular non-central $F$-distribution \cite{james}.   

The joint eigenvalue densities of non-central singular and non-singular $F$-matrix have been reported in \cite{james}. A stochastic representation of the leading eigenvalue of  a non-central $F$-matrix in the high signal-to-noise ratio (SNR) regime  is given in \cite{johnstoneRoy,prathapRoyRoot}. Certain large dimensional characteristics of the leading eigenvalue, including the phase transition phenomenon, have been analyzed in \cite{Dharmawansa2014}.
The dynamics of the eigenvalues below the phase transition have been investigated in \cite{johnstone2020}. An analysis relaxing the Gaussian assumption has been reported in a more recent dissemination \cite{hou2023spiked}. 
Notwithstanding the above facts, an exact finite dimensional statistical characterization of the leading eigenvalue of a non-central singular and non-singular $F$-matrices has not been reported in the current literature. 

Having motivated with the above facts, in this paper, capitalizing on random matrix tools, we present new exact c.d.f. expressions for the leading eigenvalue of a non-central $F$-matrix with a rank-one non-centrality parameter matrix. 
 These new c.d.f. expressions further facilitate the analysis of the receiver operating characteristics (ROC) of the largest root test. Driven by the modern high dimensional statistical applications involving the non-central $F$-matrices \cite{johnstone2020,hou2023spiked}, we have extended our analysis to the high dimensional regime in which $m,n,p\to\infty$ such that $m/p\to c_1\in(0,1)$ and $m/n\to c_2\in(0,1)$. Our key results developed here reveal that, in this regime, reliable signal detection is possible with the leading eigenvalue, provided that the signal strength is above a certain threshold (i.e., the phase transition threshold). However, those signals with strengths below this threshold cannot be detected.
 
{\it Notation}: The following notation is used throughout this paper. 
A complex Gaussian random vector $\boldsymbol{y}\in\mathbb{C}^m$ with mean $\boldsymbol{\mu}\in\mathbb{C}^{m}$ and positive definite covariance matrix $\boldsymbol{\Sigma}\in\mathbb{C}^{m\times m}$ is denoted by $\boldsymbol{y}\sim \mathcal{CN}_m(\boldsymbol{\mu},\boldsymbol{\Sigma})$. 
The superscript $(\cdot)^\dagger$ indicates the Hermitian transpose, and $\text{tr}(\cdot)$ represents the trace of a square matrix. If $y_k\sim \mathcal{CN}_m(\boldsymbol{\mu}_k,\boldsymbol{\Sigma}),k=1,2,\ldots,N$, are independent, then $\sum_{k=1}^N \boldsymbol{y}_k\boldsymbol{y}_k^\dagger$ is said to follow a complex non-central Wishart distribution denoted by $\mathcal{CW}_m\left(N,\boldsymbol{\Sigma},\boldsymbol{\Omega}\right)$, where $\boldsymbol{\Omega}=\boldsymbol{\Sigma}^{-1}\boldsymbol{M}\boldsymbol{M}^\dagger$ with $\boldsymbol{M}=\left(\boldsymbol{\mu}_1\;\ldots\;\boldsymbol{\mu}_N\right)\in\mathbb{C}^{m\times N}$ is the non-centrality parameter. 
The $m\times m$ identity matrix is represented by $\boldsymbol{I}_m$. A diagonal matrix with the diagonal entries $a_1,a_2,\ldots, a_n$ is denoted by $\text{diag}(a_1, a_2,\ldots,a_n)$. The determinant of an $n\times n$ matrix with its  $i,j$th entry given by $a_{i,j}$ is represented by $\det\left[a_{i,j}\right]_{i,j=1,\ldots,n}$  . Finally, we use the following notation to compactly represent the
determinant of an $n\times n$ block matrix:
\begin{equation*}
\begin{split}
\det\left[a_{i,j}\;\; b_{i}\right]_{\substack{i=1,2,\ldots,n\\
j=1,2,\ldots,n-1}}&=\left|\begin{array}{ccccc}
 a_{1,1} & a_{1,2}& a_{1,3}& \ldots & b_{1}\\
  \vdots & \vdots & \vdots &\ddots & \vdots \\
  a_{n,1} & a_{n,2}& a_{n,3}& \ldots & b_{n}
 \end{array}\right|.
 \end{split}
\end{equation*}

\section{Detection Problem formulation}
 	Consider the general linear signal observation model:
  \begin{align}  \boldsymbol{x}_i=\boldsymbol{A}\boldsymbol{s}+\boldsymbol{n}_i,\;\;i=1,2,\ldots,p
  \end{align}
	where $p$ is the number of observations (samples), $\boldsymbol{x}_i\in\mathbb{C}^{m}$, $\boldsymbol{A}\in\mathbb{C}^{m\times k}$ is an unknown non-random matrix, $\mathbf{s}\in\mathbb{C}^{k}$ is the unknown signal, and  $\boldsymbol{n}_i\sim \mathcal{CN}_m(\mathbf{0}, \boldsymbol{\Sigma})$ denotes the colored noise which is independent of $\boldsymbol{s}$. Furthermore, the noise covariance matrix $\boldsymbol{\Sigma}$ is also unknown to the detector. 
 \begin{rk}\label{rk1}
     It is worth noting that the vector $\boldsymbol{As}$ degenerates into simple structure for various practically important sensing applications. For instance, it takes the form $\boldsymbol{a}\boldsymbol{b}^T \boldsymbol{s}$, where $\boldsymbol{a}\in\mathbb{C}^m$ and $\boldsymbol{b}\in\mathbb{C}^{k}$ with $\max\left\{\vert\vert \boldsymbol{a}\vert \vert,\vert\vert \boldsymbol{b}\vert \vert,\vert\vert \boldsymbol{s}\vert \vert \right\}<K$, for MIMO radar applications \cite{li2007mimo,haimovich2007mimo}, whereas it assumes $\boldsymbol{a}s$, where $s\in\mathbb{C}$, for other conventional radar related applications \cite{roman2000parametric,smith2023exploiting,de2007rao}.
 \end{rk}  
 
 Consequently, the classical signal detection problem reduces to the following binary hypothesis testing problem
	\begin{align*}
	&\mathcal{H}_0:\; \boldsymbol{s}=\boldsymbol{0}\;\;\;\;\;\; \text{Signal is absent}\\
	& \mathcal{H}_1:\; \boldsymbol{s}\neq \boldsymbol{0} \;\;\;\;\; \text{Signal is present}.
	\end{align*}
 If the noise covariance matrix $\boldsymbol{\Sigma}$ was known in advance, then the noise whitening operation would result in
 \begin{align}
\boldsymbol{\Sigma}^{-1/2}\boldsymbol{x}_i=\boldsymbol{\Sigma}^{-1/2}\boldsymbol{A}\boldsymbol{s}+ \boldsymbol{w}_i
 \end{align}
where $\boldsymbol{w}_i=\boldsymbol{\Sigma}^{-1/2}\boldsymbol{n}_i\sim\mathcal{CN}_m\left(\boldsymbol{0},\boldsymbol{I}_m\right)$ denotes the white Gaussian noise. Against this backdrop, in the absence of detailed knowledge about non-random $\boldsymbol{A}$ and $\boldsymbol{s}$, certain group invariant arguments demonstrate that generic tests  depend on the eigenvalues of $\boldsymbol{\Sigma}^{-1/2}\boldsymbol{\hat{R}}\boldsymbol{\Sigma}^{-1/2}$ \cite{johnstoneRoy,muirhead}, where
\begin{align}
\boldsymbol{\hat{R}}=\frac{1}{p}\sum_{j=1}^p \boldsymbol{x}_j\boldsymbol{x}_j^\dagger
\end{align}
denotes the sample covariance matrix. Although conventionally we assume that $p\geq m$ to guarantee the almost sure positive definiteness of $\boldsymbol{\hat{R}}$ \cite{muirhead}, here we do not impose that condition as we allow it to be positive semi-definite in general. For instance, as delineated in \cite{vallet2017performance,Vinogradova}, such sample deficiency (i.e., $p<m$) is becoming increasingly common in the modern array processing due to the increased antenna size and the urge for faster decision making. Nevertheless, this in turn will make the matrix $\boldsymbol{\Sigma}^{-1/2}\boldsymbol{\hat{R}}\boldsymbol{\Sigma}^{-1/2}$ rank deficient and thereby singular.
\begin{rk}
    It is noteworthy that we have not centered the sample covariance matrix with the estimated sample mean of the unknown vector $\boldsymbol{As}$, since such a centering operation would not change the distributional characteristics other than  the number of degrees of freedom.
\end{rk}

Since $\boldsymbol{\Sigma}$ is not-known at the receiver, it is customary to replace it with the sample estimate given by
\begin{align}
    \widehat{\boldsymbol{\Sigma}}=\frac{1}{n}\sum_{\ell=1}^n \boldsymbol{n}_\ell\boldsymbol{n}_\ell^\dagger,\;\;\;\;\; n\geq m
\end{align}
where the condition $n\geq m$ guarantees the almost sure invertibility of the sample covariance matrix. As such, the parameter $n-m$ is an explicit indicator of the quality of the sample covariance estimate. In this regard, $n-m=0$ or $n=m$ configuration gives the worst possible, and yet invertible sample covariance.
To facilitate the computing of $\widehat{\boldsymbol{\Sigma}}$ , we need to make the common assumption that the availability of noise-only (i.e., signal free) samples \cite{rong2022adaptive,richmond,zachariah,dogandzic2003generalized,hemimo,chamain2020eigenvalue,nadakuditi} $\boldsymbol{n}_1,\boldsymbol{n}_2,\ldots,\boldsymbol{n}_n$ at the detector. Therefore, we can conveniently focus on a test based on the eigenvalues of $\widehat{\boldsymbol{\Sigma}}^{-1/2}\boldsymbol{\widehat R}\widehat{\boldsymbol{\Sigma}}^{-1/2}$. 

In this respect, the leading sample eigenvalue (a.k.a. Roy's largest root) has been popular among the detection theorists \cite{Kritchman,YCliang,prathapRoyRoot,larson,wang2017stat}. This is further highlighted, as delineated in \cite{johnstoneRoy}, by the fact that the Roy's largest root test is most powerful among the common tests when the alternative is of rank-one (a.k.a. the concentrated non-centrality) \cite{Kritchman}. Therefore, in light of the above discussion, we choose
$\lambda_{\max}\left(\widehat{\boldsymbol{\Sigma}}^{-1/2}\boldsymbol{\widehat R}\widehat{\boldsymbol{\Sigma}}^{-1/2}\right)=\lambda_{\max}\left(\widehat{\boldsymbol{\Sigma}}^{-1}\boldsymbol{\widehat R}\right)$
as the test statistic. Now noting that $\boldsymbol{x}_j\sim \mathcal{CN}_m\left(\boldsymbol{Ap},\boldsymbol{\Sigma}\right)$, we obtain
\begin{align*}
	&\mathcal{H}_0:\; n\widehat{\boldsymbol{\Sigma}}\sim \mathcal{CW}_m\left(n,\boldsymbol{\Sigma}, \boldsymbol{0}\right)\;\;\&\;\; p\widehat{\boldsymbol{R}}\sim \mathcal{CW}_m\left(p,\boldsymbol{\Sigma},\boldsymbol{0}\right)\\
	& \mathcal{H}_1:\; n\widehat{\boldsymbol{\Sigma}}\sim \mathcal{CW}_m\left(n,\boldsymbol{\Sigma}, \boldsymbol{0}\right)\;\;\&\;\; p\widehat{\boldsymbol{R}}\sim \mathcal{CW}_m\left(p,\boldsymbol{\Sigma},\boldsymbol{\Omega}\right)
\end{align*}
where $\boldsymbol{\Omega}=p\boldsymbol{\Sigma}^{-1}\boldsymbol{As}\boldsymbol{s}^\dagger\boldsymbol{A}^\dagger$ is the rank-one non-centrality parameter matrix. Under the above Gaussian setting,  the matrix $\widehat{\boldsymbol{\Sigma}}^{-1}\boldsymbol{\widehat R}$ is referred to as the $F$-matrix \cite{johnstoneRoy,wang2017stat,muirhead}. In particular, under $\mathcal{H}_0$ it is called a central $F$-matrix, whereas under the alternative it becomes a non-central $F$-matrix. Moreover, for $p<m$, they are referred to as singular central $F$-matrix and singular non-central $F$-matrix, respectively.

For future use, let us denote the maximum eigenvalue of $\lambda_{\max}\left(\widehat{\boldsymbol{\Sigma}}^{-1}\boldsymbol{\widehat R}\right)$ as $\hat{\lambda}_{\max}$. Now the  test based on the leading eigenvalue detects a signal if $\hat{\lambda}_{\max}>\xi_{\text{th}}$, where $\xi_{\text{th}}$ is the threshold corresponding to a desired false alarm rate $\alpha\in(0,1)$ given by
\begin{align} 
\label{pf}
\alpha=P_F(\xi_{\text{th}})=\Pr\left(\hat{\lambda}_{\max}>\xi_{\text{th}}|\mathcal{H}_0\right). 
\end{align}
Consequently, the probability of detection becomes
\begin{align}
\label{pd}
    P_D(\boldsymbol{\Omega}, \xi_{\text{th}})=\Pr\left(\hat{\lambda}_{\max}>\xi_{\text{th}}|\mathcal{H}_1\right). 
\end{align}
Subsequent elimination of $\xi_{\text{th}}$ yields a functional relationship between $P_D$ and $P_F$, which is referred to as the ROC profile.

The above computational machinery relies on the availability of the c.d.f.s of $\hat{\lambda}_{\max}$ under both hypotheses. To this end, we need to statistically characterize the c.d.f. of the maximum eigenvalue of  $\widehat{\boldsymbol{\Sigma}}^{-1}\boldsymbol{\widehat R}$. 
Therefore, in what follows, we take advantage of random matrix theory tools to derive the exact c.d.f. of the $\hat{\lambda}_{\max}$ under the alternative and thereby characterizing the ROC profile of the detector. 

\section{C.D.F of the Maximum Eigenvalue}

Let us recall the constituent distributions under the alternative  to yield
\begin{align}
    n\widehat{\boldsymbol{\Sigma}}\sim \mathcal{CW}_m\left(n,\boldsymbol{\Sigma}, \boldsymbol{0}\right)\;\;\&\;\; p\widehat{\boldsymbol{R}}\sim \mathcal{CW}_m\left(p,\boldsymbol{\Sigma},\boldsymbol{\Omega}\right)
\end{align}
where $\boldsymbol{\Omega}=p\boldsymbol{\Sigma}^{-1}\boldsymbol{As}\boldsymbol{s}^\dagger\boldsymbol{A}^\dagger$ is a rank-one matrix. Now
capitalizing on the fact that the eigenvalues of $\widehat{\boldsymbol{\Sigma}}^{-1}\boldsymbol{\widehat R}$ do not change under the transformations $\widehat{\boldsymbol{\Sigma}}\mapsto \boldsymbol{\Sigma}^{-1/2}\widehat{\boldsymbol{\Sigma}}\boldsymbol{\Sigma}^{-1/2}$ and $\widehat{\boldsymbol{\Sigma}}\mapsto \boldsymbol{\Sigma}^{-1/2}\boldsymbol{\widehat R}\boldsymbol{\Sigma}^{-1/2}$, we may equivalently consider the c.d.f. of $\hat{\lambda}_{\max}$, under the alternative, when 
\begin{align}
    n\widehat{\boldsymbol{\Sigma}}\sim \mathcal{CW}_m\left(n,\boldsymbol{I}_m, \boldsymbol{0}\right)\;\;\&\;\; p\widehat{\boldsymbol{R}}\sim \mathcal{CW}_m\left(p,\boldsymbol{I}_m,\boldsymbol{\Omega}\right)
\end{align}
where, for convenience, we have written the non-centrality parameter in its symmetric form as $\boldsymbol{\Omega}=p\boldsymbol{\Sigma}^{-1/2}\boldsymbol{As}\boldsymbol{s}^\dagger\boldsymbol{A}^\dagger \boldsymbol{\Sigma}^{-1/2}$. Therefore, we first derive the c.d.f. of a generic complex non-central $F$-matrix with a rank-one non-centrality parameter and subsequently particularize it to the c.d.f. of $\hat{\lambda}_{\max}$. The case corresponding to the singular $F$-matrix follows in the same way.

Let $\boldsymbol{S}\sim\mathcal{CW}_m\left(n,\boldsymbol{I}_m\right)$ and $\boldsymbol{R}\sim\mathcal{CW}_m\left(p,\boldsymbol{I}_m,\boldsymbol{\Theta\Theta}^H\right)$, where $\boldsymbol{\Theta\Theta}^H\in\mathbb{C}^{m\times m}$ denotes the non-centrality parameter, be independently distributed.
Now following \cite{james}, the joint density of the eigenvalues, $\lambda_1<\lambda_2<\cdots<\lambda_m$, of non-central-$F$ matrix $\boldsymbol{S}^{-1}\boldsymbol{R}$ can be written as
\begin{align*}
f(\lambda_1,\ldots,\lambda_m)&=
    K_m \prod_{k=1}^m \frac{\lambda_j^{p-m}}{\left(1+\lambda_k\right)^{p+m}}
\Delta^2_m(\boldsymbol{\lambda})\nonumber\\
    &\qquad   \times 
{}_1\widetilde{F}_1\left(n+p;p;\boldsymbol{\Lambda}_{\boldsymbol{\theta}},\left(\boldsymbol{I}_m+\boldsymbol{\Lambda}^{-1}\right)^{-1}\right)
\end{align*}
where $\Delta_m(\boldsymbol{\lambda})=\prod_{1\leq i<j\leq m}\left(\lambda_j-\lambda_i\right)$ is the Vandermonde determinant,  ${}_1\widetilde{F}_1(\cdot;\cdot;\cdot,\cdot)$ denotes the confluent hypergeometric function of two matrix arguments \cite{james},
$\boldsymbol{\Lambda}=\text{diag}\left(\lambda_1,\lambda_2,\cdots,\lambda_m\right)$, 
$\boldsymbol{\Lambda}_{\boldsymbol{\theta}}=\text{diag}\left(\theta_1,\theta_2,\cdots,\theta_m\right)$ with $\theta_1<\theta_2<\cdots<\theta_m$ denoting the eigenvalues of $\boldsymbol{\Theta\Theta}^H$, and $K_m=e^{-\text{tr}\left(\boldsymbol{\Theta\Theta}^H\right)}\prod_{k=1}^m \frac{(p+n-k)!}{(p-k)!(n-k)!(m-k)!}.$
An alternative representation of the above ${}_1\widetilde{F}_1$ function due to Khatri \cite{khatri} is given by
\begin{align}
  &{}_1\widetilde{F}_1\left(n+p;p;\boldsymbol{\Lambda}_{\boldsymbol{\theta}},\left(\boldsymbol{I}_m+\boldsymbol{\Lambda}^{-1}\right)^{-1}\right) \nonumber\\
  &= C_m \frac{\text{det}\left[{}_1 F_1\left(n+\beta+1;\beta+1;\frac{{\lambda_i}\theta_j}{1+\lambda_i}\right)\right]_{i,j=1,\ldots,m}}
  { \Delta_m(\boldsymbol{\theta})\Delta_m(\boldsymbol{\lambda})\prod_{j=1}^m(1+\lambda_j)^{m-1}} 
\end{align}
where $\beta=p-m$,  ${}_1 F_1 (\cdot;\cdot;\cdot)$ is the confluent hypergeometric function \cite{gradshteyn} and 
$
    C_m=\prod_{k=1}^m \frac{(k-1)!(\beta+1)_{m-k}}{(n+\beta+1)_{m-k}}$
with $(a)_k=a(a+1)\cdots (a+k-1)$ denoting the Pochhammer symbol.

The c.d.f. of $\lambda_{\max}$ of $\boldsymbol{S}^{-1}\boldsymbol{R}$ can be written as
\begin{align}
    \Pr\left\{\lambda_{\max}\leq t\right\}=\idotsint
    \limits_{\mathcal{R}_t} 
    f(\lambda_1,\ldots,\lambda_m)
     d{\boldsymbol{\lambda}} 
\end{align}
where $\mathcal{R}_t=\left\{0<\lambda_1<\lambda_2< \cdots<\lambda_m\leq t\right\}$ and $d\boldsymbol{\lambda}=d\lambda_1\cdots d\lambda_m$. A careful inspection of the joint density reveals that it depends on $\lambda_k$ through $\lambda_k/(1+\lambda_k)$, for $k=1,2,\ldots,m$. Therefore, noting that the map $x_k\mapsto\frac{x_k}{1+x_k}$ preserves the order, we find it convenient to find the c.d.f. of the maximum of the transformed eigenvalues $x_k=\frac{\lambda_k}{1+\lambda_k},\; k=1,2,\ldots,m$, where $0<x_1<x_2<\cdots<x_m<1$. As such, we obtain the following c.d.f. relationship between the two random variables
\begin{align}
\label{eq cdf lam}
 \Pr\left\{\lambda_{\max}\leq t\right\}=\Pr\left\{x_{\max}\leq t/1+t\right\}.
\end{align}
Therefore, in what follows, we focus on the multiple integral given by
\begin{align*}
    &\Pr\left\{x_{\max}\leq x\right\}=\idotsint
    \limits_{\mathcal{R}_x} 
g(x_1,\cdots,x_m) \prod_{k=1}^m \left(1-x_k\right)^{-2} 
     d\boldsymbol{x}
\end{align*}
where $g(x_1,\cdots,x_m)=f\left(\frac{x_1}{1-x_1},\ldots,\frac{x_m}{1-x_m}\right)$ and $\mathcal{R}_x=\left\{0<x_1<x_2<\cdots<x_m\leq x\right\}$. After some algebraic manipulation, the function $g(x_1,\cdots,x_m)$ takes the form
\begin{align*}
g(x_1,\cdots,x_m)
    &=K_mC_m \frac{\Delta_m(\boldsymbol{x})}{\Delta_m(\boldsymbol{\theta})}\prod_{k=1}^m x_k^{\beta}(1-x_k)^{\alpha+2}\nonumber\\
    & \hspace{1.5cm} \times \det\left[{}_1 F_1\left(\beta_{n,1};\beta_{0,1};x_i\theta_j\right)\right]_{i,j=1,\ldots,m}
\end{align*}
where $\alpha=n-m$ and 
$\beta_{k,\ell}=\beta+k+\ell$. Noting that $\prod_{1\leq i<j\leq m}\left(x_j-x_i\right)=\det\left[x_i^{j-1}\right]_{i,j=1,\cdots,m}$, we may use \cite[Corollary 2]{chiani} to evaluate the above multiple integral as
\begin{align}
\label{eq cdf different}
    \Pr\left\{x_{\max}\leq x\right\}=
    K_mC_m \frac{\det\left[\phi_{i}(x,\theta_j)\right]_{i,j=1,\ldots,m}}
    {\Delta_m(\boldsymbol{\theta})}
\end{align}
where $\phi_{i}(x,\theta_j)=\int_0^x y^{\beta_{0,i}-1} (1-y)^\alpha {}_1 F_1\left(\beta_{n,1};\beta_{0,1};\theta_j y\right) dy$.
Now a simple variable transformation yields
\begin{align}
  &\phi_{i}(x,\theta_j)\nonumber\\
  &=x^{\beta_{0,i}} \int_0^1 y^{\beta_{0,i}-1} (1-xy)^\alpha  {}_1 F_1\left(\beta_{n,1};\beta_{0,1};x\theta_j y\right) dy.
\end{align}
Although the above integral can be solved in closed-form, we find it convenient to leave it as it is and focus on the degenerated solution of the c.d.f. corresponding to the rank-one non-centrality parameter, which is the case of our interest.


In the above analysis, we have assumed that the $\theta_k,\; k=1,2,\ldots,m$, are distinct parameters. However, we are interested in the case for which $\boldsymbol{\Theta \Theta}^H$ is rank-one. This in turn implies $
    \theta_k=\left\{\begin{array}{ll}
    0 & k=1,2,\ldots,m-1,\\
    \theta & k=m
    \end{array}\right.$, 
where $\theta=\text{tr}\left(\boldsymbol{\Theta \Theta}^H\right)$.
The direct substitution of the above values into (\ref{eq cdf different}) would yield $0/0$ indeterminate form. To circumvent this difficulty, we evaluate the following limit denoted by   
\begin{align}
L(\theta)=\lim_{\theta_{1},\cdots,\theta_{m-1}\to 0}
\frac{\det\left[\phi_{i}(x,\theta_j)\;\;\;\; \phi_{i}(x,\theta)\right]_{\substack{i=1,\ldots,m\\j=1,\ldots,m-1}}}
    {\det\left[\theta_j^{i-1}\;\;\;\; \theta^{i-1}\right]_{\substack{i=1,\ldots,m\\j=1,\ldots,m-1}}},
\end{align}
which in light of \cite{khatri} takes the form
\begin{align*}
\frac{\displaystyle \lim_{\theta_{1},\cdots,\theta_{m-1}\to 0}\det\left[D^{m-j-1}\phi_{i}(x,\theta_j)\;\;\;\; \phi_{i}(x,\theta)\right]_{\substack{i=1,\ldots,m\\j=1,\ldots,m-1}}}
    { \displaystyle \lim_{\theta_{1},\cdots,\theta_{m-1}\to 0} \det\left[D^{m-j-1}\theta_j^{i-1}\;\;\;\; \theta^{i-1}\right]_{\substack{i=1,\ldots,m\\j=1,\ldots,m-1}}}.
\end{align*}
where $D^\ell_k\equiv \frac{d}{d\theta^\ell _k} (\cdot)$, $k,\ell=0,1,\cdots$. Since we can show, after some algebraic manipulation, that, for $j=1,2,\cdots,m-1$,
\begin{align*}
\lim_{\theta_j\to 0}D^{m-j-1}\phi_{i}(x,\theta_j)
 &= \frac{x^{p+i-j-1}(\beta_{n,1})_{m-j-1}}{(p+i-j-1)(\beta_{0,1})_{m-j-1}}\nonumber\\
&\qquad \times {}_2 F_1\left(-\alpha,p+i-j-1;p+i-j;x\right),
\end{align*}
where ${}_2 F_1 (\cdot,\cdot;\cdot;\cdot)$ is the Gauss hypergeometric function \cite{gradshteyn}, and $\lim_{\theta_1,\ldots,\theta_{m-1}\to 0}\det\left[D^{m-j-1}\theta_j^{i-1}\;\;\;\; \theta^{i-1}\right]_{\substack{i=1,\ldots,m\\j=1,\ldots,m-1}} = (-1)^{\lfloor \frac{m-1}{2} \rfloor}\prod_{k=1}^{m-2}k! \theta^{m-1}$,
where $\lfloor \cdot \rfloor$ is the floor function, we readily obtain
\begin{align}
L(\theta)&=(m-1)!x^{(p-1)m+1}\theta^{1-m}\prod_{k=1}^{m-1}
    \frac{(\beta_{n,1})_{m-k-1}}{k!(\beta_{0,1})_{m-k-1}}\nonumber\\
    &  \qquad \qquad \qquad \times 
    \det\left[\psi_{i,j}(x)\;\;\;\; \xi_i(\theta, x)\right]_{\substack{i=1,\ldots,m\\j=1,\ldots,m-1}}
\end{align}
where $\psi_{i,j}(x)={}_2 F_1\left(-\alpha,p+i-j-1;p+i-j;x\right)/(p+i+j-m-1)$
    and $\xi_i(\theta, x)=\sum_{k=0}^\alpha 
    \frac{(-x)^k {}_2 F_2\left(\beta_{n,1},\beta_{i,k};\beta_{0,1},\beta_{i+1,k};\theta x\right)}{k!(\alpha-k)!\beta_{i,k}}$
with ${}_2 F_2 (\cdot,\cdot;\cdot,\cdot;\cdot)$ denoting the generalized hypergeometric function \cite{gradshteyn}.
Consequently, we use the above expression in (\ref{eq cdf different}) with some algebraic manipulation to obtain the c.d.f. as
\begin{align}
    \Pr\left\{x_{\max}\leq x\right\}&=
\mathcal{K}_{m,n}\theta^{1-m}e^{-\theta} x^{(p-1)m+1}\nonumber\\
    &\quad \times 
    \det\left[\psi_{i,j}(x)\;\; \xi_i(\theta, x)\right]_{\substack{i=1,\ldots,m\\j=1,\ldots,m-1}}
\end{align}
where $\mathcal{K}_{m,n}=\frac{(n+\beta)!}{\beta!}\prod_{k=1}^{m-1}
    \frac{(p+n-k-1)!}{(n-k)!(m-k-1)!(p-k-1)!}$.
Finally, in light of (\ref{eq cdf lam}), we obtain the c.d.f. of $\lambda_{\max}$, which is given by the following theorem. 
\begin{thm}
\label{thm cdf}
    The c.d.f. of the maximum eigenvalue of $\boldsymbol{S}^{-1}\boldsymbol{R}$, for rank-one non-centrality parameter, is given by
    \begin{align*}
    F_{\lambda_{\max}}(t;\theta)&=
    \mathcal{K}_{m,n}\theta^{1-m}e^{-\theta} \left(\frac{t}{1+t}\right)^{(p-1)m+1}\nonumber\\
    & \times 
\det\left[\psi_{i,j}\left(\frac{t}{1+t}\right)\;\; \xi_i\left(\theta, \frac{t}{1+t}\right)\right]_{\substack{i=1,\ldots,m\\j=1,\ldots,m-1}}.
\end{align*}
\end{thm}
Notwithstanding that, for $p<m$, $\widehat{\boldsymbol{R}}$ does not have a density, we may utilize the joint density of the non-zero eigenvalues of singular $F$-matrix given by \cite{james} to arrive at the corresponding c.d.f. of $\lambda_{\max}$ as shown in the following corollary.
\begin{cor}\label{cor1}
    The c.d.f. of the maximum eigenvalue of {\it singular} $\boldsymbol{S}^{-1}\boldsymbol{R}$, for rank-one non-centrality parameter, is obtained from $F_{\lambda_{\max}}(t;\theta)$ by replacing $m$ by $p$, $p$ by $m$, and $n$ by $n+p-m$.
\end{cor}

Having derived the finite dimensional c.d.f. of the leading eigenvalue of a non-central $F$-matrix, we are now ready to analyze the ROC profiles of the proposed test.

\section{ROC of the Largest Eigenvalue Based Test }
Here we analyze the ROC performance associated with the maximum eigenvalue based test for finite as well as asymptotic regimes. 
To this end, by exploiting the relationship between the eigenvalues  of $\widehat{\boldsymbol{\Sigma}}^{-1}\boldsymbol{\widehat R}$ and that of $\boldsymbol{S}^{-1}\boldsymbol{R}$ given by by $\hat{\lambda}_j=(n/p)\lambda_j$, for $j=1,2,\ldots,m$, we may express the c.d.f. of the maximum eigenvalue of $\widehat{\boldsymbol{\Sigma}}^{-1}\boldsymbol{\widehat R}$ as
    $F_{\lambda_{\max}}(\kappa x;\gamma)$, where $\kappa=p/n$ and $\gamma=\text{tr}\left(\boldsymbol{\Omega}\right)=p\boldsymbol{s}^\dagger \boldsymbol{A}^\dagger \boldsymbol{\Sigma}^{-1}\boldsymbol{As}$. 
    
    \subsection{Finite Dimensional Analysis}
    Therefore, in light of Theorem \ref{thm cdf} and (\ref{pd}), we get
    \begin{align}
        P_D(\gamma, \xi_{\text{th}})=1-F_{\lambda_{\max}}(\kappa \xi_{\text{th}};\gamma).
    \end{align}
Now following \cite[Corollary 8]{chamain2020eigenvalue}, we obtain the false alarm probability as
\begin{align}
    P_F(\xi_{\text{th}})=1-F_{\lambda_{\max}}(\kappa \xi_{\text{th}})
\end{align}
where 
\begin{align}
			F_{\lambda_{\max}}(t)
			&=\mathcal{K}\dfrac{(n+p-1)!}{(m+p-1)!}\left(\dfrac{t}{1+t}\right)^{m(n+p-m)}\nonumber\\
			& \qquad \qquad \quad \times \det\left[\Psi_{i+1,j+1}(t)\right]_{i,j=1,...,\alpha}.
		\end{align}
  with $\Psi_{i,j}(t)= (p+i-1)_{j-2}P_{m+i-j}^{(j-2,\beta+j-2)}\left(\frac{2}{t}+1\right)$, and
$\mathcal{K}=\prod_{j=0}^{\alpha-1}\dfrac{(p+m+j-1)!}{(p+m+2j)!}$. Here $P_c^{(a,b)}(z)$ denotes the Jacobi polynomial\cite{gradshteyn}. Again, corresponding expressions related to $p<m$ scenario can easily be obtained by capitalizing on the parameter transformations given in Corollary \ref{cor1}. We do not mention them here due to space limitations.
  
  Obtaining a function relationship between $P_D$ and $P_F$ (i.e., ROC profile) by eliminating the dependency on $\xi_{\text{th}}$ seems to be an arduous task for a general parameter configuration of $m,n$ and $p$. However, such an explicit relationship is feasible for the important configuration of $n=m$. To be specific, for $m=n$, we get $P_F(\xi_{\text{th}})=1-(\kappa \xi_{\text{th}}/1+\kappa\xi_{\text{th}})^{mp}$, dwelling on which we obtain the corresponding ROC profile as
  \begin{align}
  \label{rocclosed}
      P_D&=1-\mathcal{K}_{m,m}\gamma^{1-m}e^{-\gamma}\left(1-P_F\right)^{1-\frac{1}{p}\left(1-\frac{1}{m}\right)}\nonumber\\
      & \quad \times \det\left[\frac{1}{\beta_{i,j}-1}\;\; \eta_i\left(\gamma\left[1-P_F\right]^{\frac{1}{mp}}\right)\right]_{\substack{i=1,\ldots,m\\j=1,\ldots,m-1}}
  \end{align}
  where $\eta_i(z)=\left(\beta_{i,0}\right)^{-1}
    {}_2 F_2\left(\beta_{n,1},\beta_{i,0};\beta_{0,1},\beta_{i,1}; z\right)$.
The ROC profile, for $p<m$, can be obtained from (\ref{rocclosed}) by  using the parameter transformations given in Corollary \ref{cor1}.
\begin{figure}[t!]
    \centering
\includegraphics[width=0.4\textwidth]{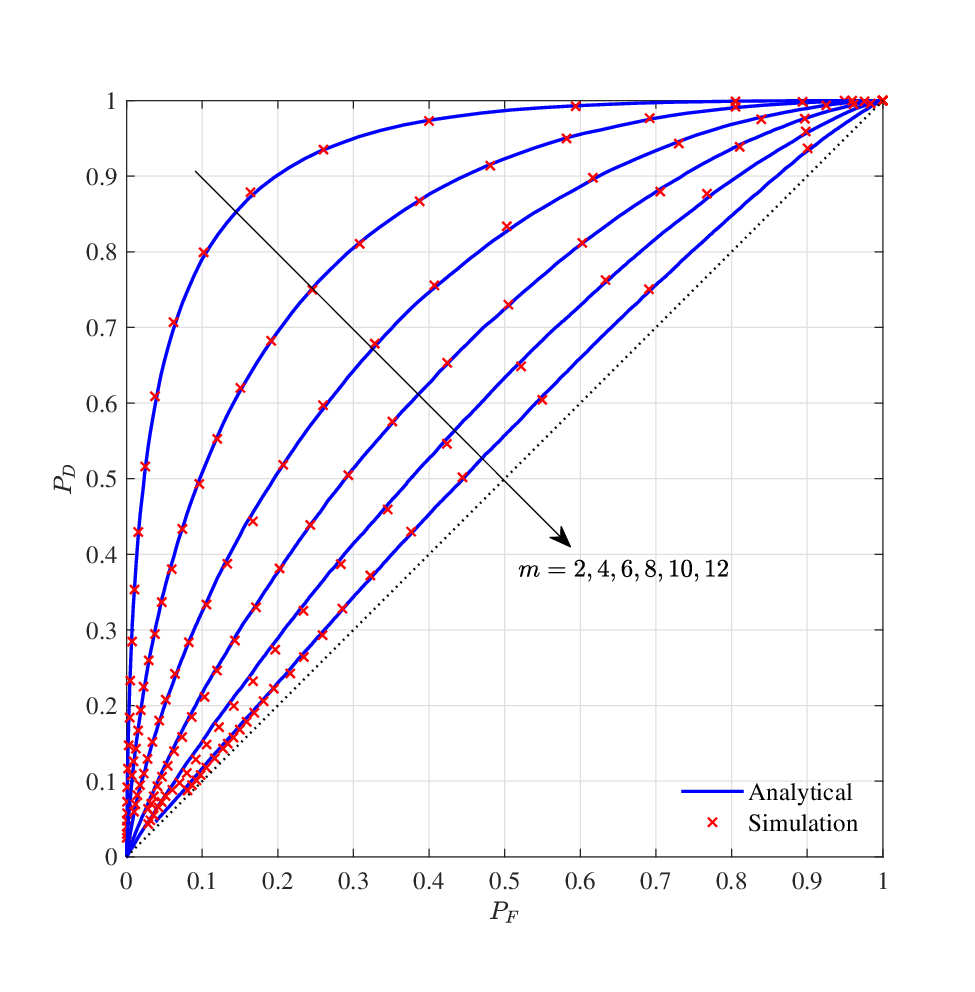}
\vspace{-7mm}
    \caption{The effect of $m$ on the ROC for $n=12,\;p=10$, and $\gamma=20$.}
    \vspace{-3mm}
    \label{fig:m}
\end{figure}
The effect of $m$ on the ROC profiles is depicted in fig. \ref{fig:m}, where $p=10,\; n=12$, and $\gamma=20$. Here $m=12$ corresponds to the singular $F$- distribution. As can be seen from the figure, disparity between $p$ and $m$ (i.e., $p-m$) improves the detector performance. The same observation is valid for $n$ and $m$ as well. Clearly $m=n$ configuration dictates a performance lower bound on the detector for fixed $p$ and $\gamma$.

\subsection{High Dimensional Analysis}
Many modern detectors are of high dimensional in nature \cite{nadakuditi,vallet2017performance,Vinogradova}. In this respect, the analytical high dimensional regimes of interest include large $m,n$, and $p$ domains. In particular, we are interest in the asymptotic regime where $m,n$, and $p$ diverge to infinity so that $m/p\to c_1\in(0,1)$ and $m/n\to c_2\in(0,1)$. For tractability, in what follows, we focus on a detection scenario delineated in Remark \ref{rk1} and assume $\lambda_{\max} (\boldsymbol{\Sigma})=O(1)$ \cite{nadakuditi,bai1998no,bai1999exact} so that $\gamma=O(p)$. Notwithstanding that the c.d.f. expressions we have derived previously are important in the finite dimensional regime, their utility in the above high dimensional regime is severely restricted due to their determinantal structure. To circumvent this difficulty, capitalizing on the large dimensional random matrix theory of central $F$-matrices \cite{Johnstone2008,jiang2022invariance,han2016tracy}, as $m,n,p\to\infty$ such that $m/p\to c_1$ and $m/n\to c_2$, we observe that 
\begin{align}
\label{tracy}
    & \mathcal{H}_0:\;\; 
\hat{\lambda}_{\max}\approx \mu + m^{-2/3} \sigma_0 {\mathcal {TW}}_2
\end{align}
where $\mu=\left(\frac{1+r}{1-c_2}\right)^2$ with $r=\sqrt{c_1+c_2-c_1c_2}$, $\sigma_0^3=c_1^4(c_1+r)^4(c_1+c_2)^4/r \left((c_1+c_2)^2-c_2(c_1+r)^2\right)^4$ are non-random fixed parameters independent of $m,n,p$, and $\mathcal{TW}_2$ denotes the {\it unitary} Tracy-Widom distributed random variable \cite{tracy1994}.
However, as explained in \cite{Dharmawansa2014,johnstone2020}, since $\gamma=O(p)$, under the alternative, $\hat{\lambda}_{\max}$ may have two different stochastic representations depending on the magnitude of $\bar{\gamma}=\boldsymbol{s}^\dagger \boldsymbol{A}^\dagger \boldsymbol{\Sigma}^{-1}\boldsymbol{As}$ relative to  the phase transition threshold. To be precise, we have from \cite{Dharmawansa2014}
\begin{align}
    \mathcal{H}_1: \hat{\lambda}_{\max}\approx \left\{\begin{array}{ll}\nu + m^{-1/2} \sigma_1 {Z} & \text{if $\bar{\gamma}>\bar{\gamma}_{\text{p}}$}\\
    \mu + m^{-2/3} \sigma_0 {\mathcal {TW}}_2 & \text{if $\bar{\gamma}<\bar{\gamma}_{\text{p}}$}
    \end{array}\right.
\end{align}
where $\sigma_1^2=\frac{t^2 \bar{\gamma}^2 (1+\bar{\gamma})^2\left(\bar{\gamma}^2-c_2(1+\bar{\gamma})^2-c_1\right) }{\left(c_2-\bar{\gamma}+c_2 \bar{\gamma}\right)^4}$ with $
t^2 = c_1+c_2-\frac{c_1(\bar{\gamma}^2-c_1)}{(1+\bar{\gamma})^2}$, $\nu=(\bar{\gamma}+c_1)(1+\bar{\gamma})/(\bar{\gamma}-(1+\bar{\gamma})c_2)>\mu$, $\bar{\gamma}_{\text{p}}=(c_2+r)/(1-c_2)$ is the phase transition threshold, and $Z$ is a standard normal random variable. The above fact reveals that, in the high dimensional regime, the leading eigenvalue has no detection power to detect signals whose strengths fall below $\bar{\gamma}_{\text{p}}$ (a.k.a. weak signals). Nonetheless, strong signals can still be detected with a high precision in the high dimensional regime with the leading eigenvalue as we show below. 

\begin{figure}[t!]
    \centering
    \includegraphics[width=0.4\textwidth]{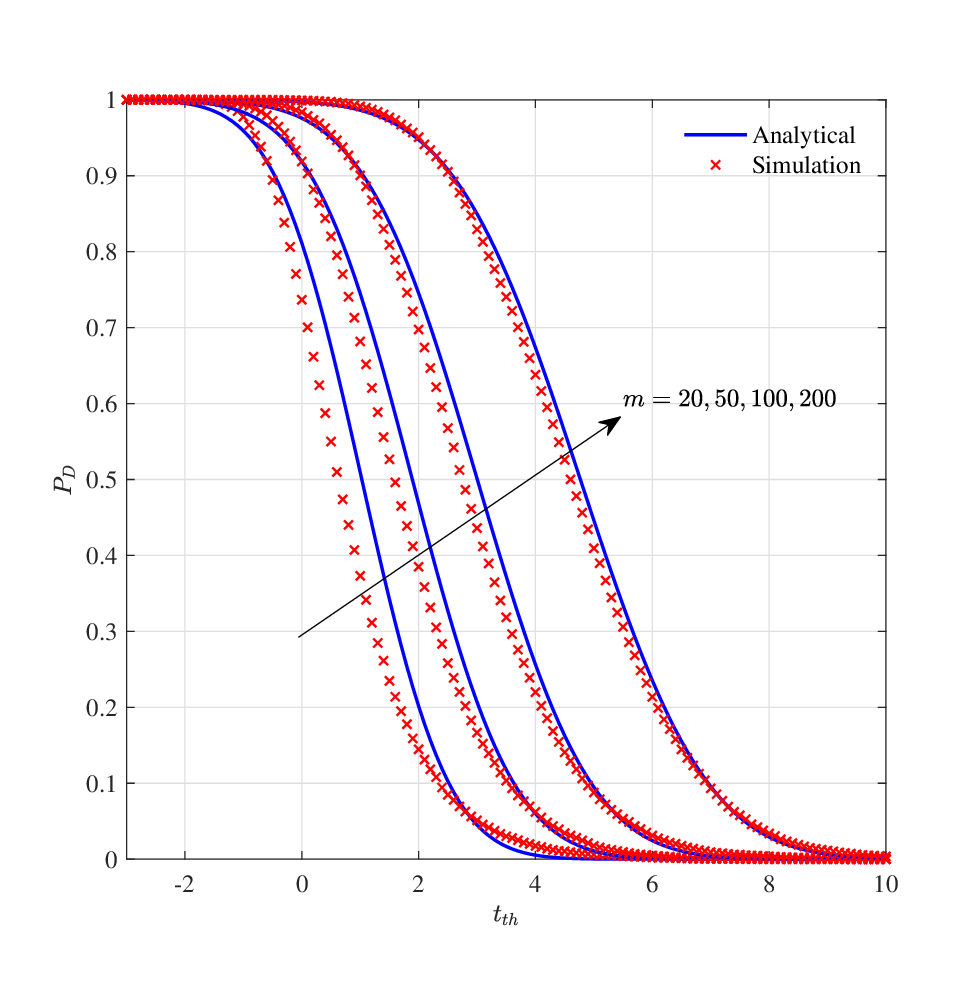}
    \vspace{-7mm}
    \caption{The effect of $m$  on $P_D$ for $c_1=0.25,\;c_2=0.5 $, and $\bar\gamma=5$. 
    }
     \vspace{-3mm}
    \label{fig:2}
\end{figure}

Assuming that $\bar{\gamma}>\bar{\gamma}_{\text{p}}$ (i.e., strong signal), let us focus on the test statistic given by
\begin{align}
\label{test stat}
t=m^{2/3}\left(\frac{\hat{\lambda}_{\max}-\mu}{\sigma_0}\right)
\end{align}
which is distributed, under the null, as the unitary Tracy-Widom as per (\ref{tracy}). Therefore, for a given fixed false alarm rate $\alpha$, we can choose a threshold $t_{\text{th}}$ such that $t_{\text{th}}=\mathcal{F}_2^{-1}(1-\alpha)$, where $\mathcal{F}_2$ denotes the unitary Tracy-Widom c.d.f. Now what remains is to show that the test static $t$ has power in this particular asymptotic domain. To this end, we rearrange the terms in (\ref{test stat}) to yield
\begin{align}
  t&\approx m^{2/3}\left(\frac{\nu-\mu+m^{-1/2} \sigma_1 Z}{\sigma_0}\right)\nonumber \\ 
  &= m^{2/3}\left(\frac{\nu-\mu}{\sigma_0}\right)+m^{1/6}\left(\frac{ \sigma_1 Z}{\sigma_0}\right).
\end{align}
Consequently, noting that $t_{\text{th}}$ does not depend on $m$, the asymptotic power of the test can readily be written as
\begin{align}
\label{power}
    P_D=\Pr\left\{t>t_{\text{th}}\right\}\approx\mathcal{Q}\left(\frac{\sigma_0 t_{\text{th}}-m^{2/3}(\nu-\mu)}{m^{1/6}\sigma_1}\right)
\end{align}
from which we obtain $P_D\to 1$ as $m\to\infty$, since $\nu>\mu$. This clearly demonstrate that $\hat{\lambda}_{\max}$ can be used to reliably detect a strong signal in the high dimensional regime. To further strengthen our claim, in fig. \ref{fig:2}, we compare $P_D$ for various $m$ values for constant $c_1$ and $c_2$. Clearly, the test statistic $t$, although optimal in the high dimensional regime, demonstrates good $P_D$ profiles for not so large system configurations as well. This further verifies the accuracy of analytical
asymptotic in  (\ref{power}).

\section{Conclusion}
This paper investigates the signal detection problem in colored noise using the leading eigenvalue of whitened signal-plus-noise sample covariance matrix. To be specific, our focus is on three regimes: sample abundance, sample  deficient, and high dimensional. In the sample abundance and deficient regimes, we have assessed the performance of this detector by developing new expressions for the c.d.f.s of the largest generalized eigenvalue of complex singular/non-singular non-central $F$-matrix with a rank-one non-centrality parameter matrix. It turns out that, in the high dimensional regime, reliable signal detection is possible if the signal strength is above a certain threshold. 




\newpage





\begin{thebibliography}{10}
\providecommand{\url}[1]{#1}
\csname url@samestyle\endcsname
\providecommand{\newblock}{\relax}
\providecommand{\bibinfo}[2]{#2}
\providecommand{\BIBentrySTDinterwordspacing}{\spaceskip=0pt\relax}
\providecommand{\BIBentryALTinterwordstretchfactor}{4}
\providecommand{\BIBentryALTinterwordspacing}{\spaceskip=\fontdimen2\font plus
\BIBentryALTinterwordstretchfactor\fontdimen3\font minus \fontdimen4\font\relax}
\providecommand{\BIBforeignlanguage}[2]{{%
\expandafter\ifx\csname l@#1\endcsname\relax
\typeout{** WARNING: IEEEtran.bst: No hyphenation pattern has been}%
\typeout{** loaded for the language `#1'. Using the pattern for}%
\typeout{** the default language instead.}%
\else
\language=\csname l@#1\endcsname
\fi
#2}}
\providecommand{\BIBdecl}{\relax}
\BIBdecl

\bibitem{nadakuditi}
R.~R. Nadakuditi and J.~W. Silverstein, ``Fundamental limit of sample generalized eigenvalue based detection of signals in noise using relatively few signal-bearing and noise-only samples,'' \emph{IEEE J. Sel. Top. Signal Process.}, vol.~4, no.~3, pp. 468--480, Jun. 2010.

\bibitem{chamain2020eigenvalue}
L.~D. Chamain, P.~Dharmawansa, S.~Atapattu, and C.~Tellambura, ``Eigenvalue-based detection of a signal in colored noise: Finite and asymptotic analyses,'' \emph{IEEE Trans. Inf. Theory}, vol.~66, no.~10, pp. 6413--6433, 2020.

\bibitem{debbah2011}
P.~Bianchi, M.~Debbah, M.~Maida, and J.~Najim, ``Performance of statistical tests for single-source detection using random matrix theory,'' \emph{IEEE Trans. Inf. Theory}, vol.~57, no.~4, pp. 2400--2419, Apr. 2011.

\bibitem{edelman}
R.~R. Nadakuditi and A.~Edelman, ``Sample eigenvalue based detection of high-dimensional signals in white noise using relatively few samples,'' \emph{IEEE Transactions on Signal Processing}, vol.~56, no.~7, pp. 2625 -- 2638, 2008.

\bibitem{mcwhorter2023}
T.~McWhorter, L.~Scharf, C.~Moore, and M.~Cheney, ``Passive multi-channel detection: A general first-order statistical theory,'' \emph{IEEE Open Journal of Signal Processing}, vol.~43, pp. 437--451, 2023.

\bibitem{johnstone2020}
I.~M. Johnstone and A.~Onatski, ``Testing in high-dimensional spiked models,'' \emph{Ann. Stat.}, vol.~44, no.~3, pp. 1231--1254, 2020.

\bibitem{YCliang}
Y.~Zeng and Y.~C. Liang, ``Eigenvalue based spectrum sensing algorithms for cognitive radio,'' \emph{IEEE Transactions on Communications}, vol.~57, no.~6, pp. 3930 -- 3941, 2009.

\bibitem{larson}
E.~Axell, G.~Leus, E.~G. Larsson, and H.~V. Poor, ``Spectrum sensing for cognitive radio: {S}tate-of-the-art and recent advances,'' \emph{IEEE Signal Process. Mag.}, vol.~29, no.~3, pp. 101--116, May 2012.

\bibitem{hemimo}
Q.~He, N.~H. Lehmann, R.~S. Blum, and A.~M. Haimovich, ``M{IMO} radar moving target detection in homogeneous clutter,'' \emph{IEEE Trans. Aerosp. Electron. Syst.}, vol.~46, no.~3, pp. 1290--1301, 2010.

\bibitem{Kritchman}
S.~Kritchman and B.~Nadler, ``Non-parametric detection of the number of signals: Hypothesis testing and random matrix theory,'' \emph{IEEE Transactions on Signal Processing}, vol.~57, no.~9, pp. 3930 -- 3941, 2009.

\bibitem{johnstoneRoy}
I.~M. Johnstone and B.~Nadler, ``{Roy's largest root test under rank-one alternatives},'' \emph{Biometrika}, vol. 104, no.~1, pp. 181--193, Mar. 2017.

\bibitem{prathapRoyRoot}
P.~Dharmawansa, B.~Nadler, and O.~Shwartz, ``Roy’s largest root under rank-one perturbations: The complex valued case and applications,'' \emph{J. Multivariate Anal.}, vol. 174, p. 104524, Nov. 2019.

\bibitem{wang2017stat}
Q.~Wang and J.~Yao, ``Extreme eigenvalues of large-dimensional spiked {F}isher matrices with application,'' \emph{Ann. Statist.}, vol.~45, no.~1, pp. 415--460, Feb. 2017.

\bibitem{johnstone2001}
I.~M. Johnstone, ``On the distribution of the largest eigenvalue in principal components analysis,'' \emph{The Annals of statistics}, vol.~29, no.~2, pp. 295--327, 2001.

\bibitem{baikPhase}
J.~Baik, G.~B. Arous, and S.~P{\'e}ch{\'e}, ``{Phase transition of the largest eigenvalue for nonnull complex sample covariance matrices},'' \emph{Ann. Probab.}, vol.~33, no.~5, pp. 1643--1697, Sep. 2005.

\bibitem{Dharmawansa2014}
P.~{Dharmawansa}, I.~M. {Johnstone}, and A.~{Onatski}, ``{Local asymptotic normality of the spectrum of high-dimensional spiked {F}-ratios},'' \emph{arXiv:1411.3875 [math.ST]}, Nov. 2014.

\bibitem{richmond}
C.~D. Richmond, ``Mean-squared error and threshold {SNR} prediction of maximum-likelihood signal parameter estimation with estimated colored noise covariances,'' \emph{IEEE Trans. Inf. Theory}, vol.~52, no.~5, pp. 2146--2164, 2006.

\bibitem{Vinogradova}
J.~Vinogradova, R.~Couillet, and W.~Hachem, ``Statistical inference in large antenna arrays under unknown noise pattern,'' \emph{IEEE Trans. Signal Process.}, vol.~61, no.~22, pp. 5633--5645, Nov. 2013.

\bibitem{werner2007}
K.~Werner and M.~Jansson, ``{DOA} estimation and detection in colored noise using additional noise-only data,'' \emph{IEEE Trans. Signal Process.}, vol.~55, no.~11, pp. 5309--5322, 2007.

\bibitem{werner2006}
------, ``Optimal utilization of signal-free samples for array processing in unknown colored noise fields,'' \emph{IEEE Trans. Signal Process.}, vol.~54, no.~10, pp. 3861--3872, 2006.

\bibitem{muirhead}
R.~J. Muirhead, \emph{Aspects of {M}ultivariate {S}tatistical {T}heory}.\hskip 1em plus 0.5em minus 0.4em\relax John Wiley \& Sons, 2009, vol. 197.

\bibitem{dogandzic2003generalized}
A.~Dogandzic and A.~Nehorai, ``Generalized multivariate analysis of variance--{A} unified framework for signal processing in correlated noise,'' \emph{IEEE Signal Process. Mag.}, vol.~20, no.~5, pp. 39--54, 2003.

\bibitem{zachariah}
D.~Zachariah, M.~Jansson, and M.~Bengtsson, ``Utilization of noise-only samples in array processing with prior knowledge,'' \emph{IEEE Trans. Signal Process. Lett.}, vol.~20, no.~9, pp. 865--868, 2013.

\bibitem{melvin2004}
W.~L. Melvin, ``A stap overview,'' \emph{IEEE Aerospace and Electronic Systems Magazine}, vol.~19, no.~1, pp. 19--35, 2004.

\bibitem{james}
A.~T. James, ``Distributions of matrix variates and latent roots derived from normal samples,'' \emph{Ann. Math. Statist.}, vol.~35, no.~2, pp. 475--501, Jun. 1964.

\bibitem{vallet2017performance}
P.~Vallet and P.~Loubaton, ``On the performance of {MUSIC} with toeplitz rectification in the context of large arrays,'' \emph{IEEE Trans. Signal Process.}, vol.~65, no.~22, pp. 5848--5859, 2017.

\bibitem{pham2015}
G.-T. Pham, P.~Loubaton, and P.~Vallet, ``Performance analysis of spatial smoothing schemes in the context of large arrays,'' \emph{IEEE Trans. Signal Process.}, vol.~64, no.~1, pp. 160--172, 2015.

\bibitem{mestre2008}
X.~Mestre and M.~{\'A}. Lagunas, ``Modified subspace algorithms for {DoA} estimation with large arrays,'' \emph{IEEE Trans. Signal Process.}, vol.~56, no.~2, pp. 598--614, 2008.

\bibitem{hou2023spiked}
Z.~Hou, X.~Zhang, Z.~Bai, and J.~Hu, ``Spiked eigenvalues of noncentral fisher matrix with applications,'' \emph{Bernoulli}, vol.~29, no.~4, pp. 3171--3197, 2023.

\bibitem{li2007mimo}
J.~Li and P.~Stoica, ``Mimo radar with colocated antennas,'' \emph{IEEE signal processing magazine}, vol.~24, no.~5, pp. 106--114, 2007.

\bibitem{haimovich2007mimo}
A.~M. Haimovich, R.~S. Blum, and L.~J. Cimini, ``Mimo radar with widely separated antennas,'' \emph{IEEE signal processing magazine}, vol.~25, no.~1, pp. 116--129, 2007.

\bibitem{roman2000parametric}
J.~R. Roman, M.~Rangaswamy, D.~W. Davis, Q.~Zhang, B.~Himed, and J.~H. Michels, ``Parametric adaptive matched filter for airborne radar applications,'' \emph{IEEE Transactions on Aerospace and Electronic Systems}, vol.~36, no.~2, pp. 677--692, 2000.

\bibitem{smith2023exploiting}
J.~Smith and A.~Shaw, ``Exploiting unit circle roots for improved radar moving target detection in low rank clutter with limited secondary data,'' \emph{IEEE Transactions on Aerospace and Electronic Systems}, 2023.

\bibitem{de2007rao}
A.~De~Maio, ``Rao test for adaptive detection in gaussian interference with unknown covariance matrix,'' \emph{IEEE transactions on signal processing}, vol.~55, no.~7, pp. 3577--3584, 2007.

\bibitem{rong2022adaptive}
Y.~Rong, A.~Aubry, A.~De~Maio, and M.~Tang, ``Adaptive radar detection in {G}aussian interference using clutter-free training data,'' \emph{IEEE Trans. Signal Process.}, vol.~70, pp. 978--993, 2022.

\bibitem{khatri}
C.~G. Khatri, ``On the moments of traces of two matrices in three situations for complex multivariate normal populations,'' \emph{Sankhy\=a}, vol.~32, no.~1, pp. 65--80, Mar. 1970.

\bibitem{gradshteyn}
I.~Gradshteyn and I.~Ryzhik, \emph{Table of Integrals, Series, and Products}, 7th~ed.\hskip 1em plus 0.5em minus 0.4em\relax Boston: Academic Press, 2007.

\bibitem{chiani}
M.~Chiani, M.~Win, and A.~Zanella, ``On the capacity of spatially correlated {MIMO} {R}ayleigh-fading channels,'' \emph{IEEE Trans. Inf. Theory}, vol.~49, no.~10, pp. 2363--2371, Oct. 2003.

\bibitem{bai1998no}
Z.-D. Bai and J.~W. Silverstein, ``No eigenvalues outside the support of the limiting spectral distribution of large-dimensional sample covariance matrices,'' \emph{The Annals of Probability}, vol.~26, no.~1, pp. 316--345, 1998.

\bibitem{bai1999exact}
Z.~Bai and J.~W. Silverstein, ``Exact separation of eigenvalues of large dimensional sample covariance matrices,'' \emph{Annals of probability}, pp. 1536--1555, 1999.

\bibitem{Johnstone2008}
I.~M. Johnstone, ``Multivariate analysis and {J}acobi ensembles: Largest eigenvalue, {T}racy-{W}idom limits and rates of convergence,'' \emph{Ann. Stat.}, vol.~36, no.~6, pp. 2638--2716, 2008.

\bibitem{jiang2022invariance}
D.~Jiang, Z.~Hou, Z.~Bai, and R.~Li, ``Invariance principle and clt for the spiked eigenvalues of large-dimensional fisher matrices and applications,'' \emph{arXiv preprint arXiv:2203.14248}, 2022.

\bibitem{han2016tracy}
X.~Han, G.~Pan, and B.~Zhang, ``The {T}racy--{W}idom law for the largest eigenvalue of {F} type matrices,'' \emph{Ann. Stat.}, vol.~44, no.~4, pp. 1564--1592, 2016.

\bibitem{tracy1994}
C.~A. Tracy and H.~Widom, ``Level-spacing distributions and the airy kernel,'' \emph{Communications in Mathematical Physics}, vol. 159, pp. 151--174, 1994.

\end{thebibliography}
\end{document}